\documentclass[%
 aip,
 amsmath,amssymb,
 reprint,%
]{revtex4-1}

\usepackage{graphicx}% Include figure files
\usepackage{dcolumn}% Align table columns on decimal point
\usepackage{bm}% bold math
\usepackage{xcolor}
\usepackage[utf8]{inputenc}
\usepackage[T1]{fontenc}
\usepackage{mathptmx}
\usepackage{etoolbox}

\makeatletter
\def\@email#1#2{%
 \endgroup
 \patchcmd{\titleblock@produce}
  {\frontmatter@RRAPformat}
  {\frontmatter@RRAPformat{\produce@RRAP{*#1\href{mailto:#2}{#2}}}\frontmatter@RRAPformat}
  {}{}
}%
\makeatother
\begin{document}

\preprint{AIP/123-QED}

\title{Self-interfering high harmonic beam arrays driven by Hermite-Gaussian beams}
% Force line breaks with \\
\author{David D. Schmidt$^\dagger$}
\email{daschmid@mines.edu}
\affiliation{Department of Physics, Colorado School of Mines, 1523 Illinois Street, Golden, CO 80401, USA}

\author{Jos\'e Miguel Pablos-Mar\'in$^{\dagger}$}
% \email{Second.Author@institution.edu.}
\affiliation{Grupo de Investigación en Aplicaciones del Láser y Fotónica, Departamento de Física Aplicada, Universidad de Salamanca, 37008, Salamanca, Spain}
\affiliation{Unidad de Excelencia en Luz y Materia Estructuradas (LUMES), Universidad de Salamanca, Salamanca, Spain}
\author{Cameron Clarke}
\author{Jonathan Barolak}
\author{Nathaniel Westlake}
%\author{David Goldberger}
\affiliation{Department of Physics, Colorado School of Mines, 1523 Illinois Street, Golden, CO 80401, USA}

\author{Alba de las Heras}
\author{Javier Serrano}
\affiliation{Grupo de Investigación en Aplicaciones del Láser y Fotónica, Departamento de Física Aplicada, Universidad de Salamanca, 37008, Salamanca, Spain}
\affiliation{Unidad de Excelencia en Luz y Materia Estructuradas (LUMES), Universidad de Salamanca, Salamanca, Spain}

\author{Sergei Shevtsov}
\author{Peter Kazansky}
\affiliation{Optoelectronics Research Centre, University of Southampton, Southampton SO17 1BJ, UK}

\author{Daniel Adams}
\affiliation{Department of Physics, Colorado School of Mines, 1523 Illinois Street, Golden, CO 80401, USA}

\author{Carlos Hern\'andez-Garc\'ia}
\affiliation{Grupo de Investigación en Aplicaciones del Láser y Fotónica, Departamento de Física Aplicada, Universidad de Salamanca, 37008, Salamanca, Spain}
\affiliation{Unidad de Excelencia en Luz y Materia Estructuradas (LUMES), Universidad de Salamanca, Salamanca, Spain}

\author{Charles G. Durfee}
\affiliation{Department of Physics, Colorado School of Mines, 1523 Illinois Street, Golden, CO 80401, USA}
\email{cdurfee@mines.edu}

% \homepage{http://www.Second.institution.edu/~Charlie.Author.}
%\affiliation{%
%Second institution and/or address%\\This line break forced% with \\
%}%

\date{\today}% It is always \today, today,

\begin{abstract}
The use of structured light to drive highly nonlinear processes in matter not only enables imprinting spatially-resolved properties onto short-wavelength radiation, but also opens alternative avenues for exploring the dynamics of nonlinear laser-matter interactions. In this work, we experimentally and theoretically explore the unique properties of driving high-order harmonic generation (HHG) with Hermite-Gaussian beams. HHG driven by Laguerre-Gauss modes results in harmonics that inherit the azimuthal Laguerre-Gauss modal structure, with their topological charge scaling according to orbital angular momentum conservation. In contrast, when HHG is driven by Hermite-Gauss beams, the harmonic modes do not show a direct correspondence to the driving modal profile.  Our experimental measurements using HG$_{0,1}$ and HG$_{1,1}$ modes, which are in excellent agreement with our numerical simulations, show that the lobes of the Hermite-Gauss driving beams effectively produce a set of separate phase-locked harmonic beamlets which can interfere downstream. This self-interference, which can be adjusted through the relative position between the gas target and the driving beam focus, can be exploited for precision extreme-ultraviolet interferometry. We demonstrate a simple application to calibrate the dispersion of an extreme-ultraviolet diffraction grating. In addition, we show through simulations that the array of harmonic beamlets can be used as an illumination source for single-shot extreme-ultraviolet ptychography. 
\end{abstract}

\maketitle

$\dagger$ These authors contributed equally to this work

\section{\label{sec:intro1}Introduction}

High harmonic generation (HHG) has been a subject of continuing interest as a means to generate coherent light in the extreme ultraviolet (EUV) for applications in spectroscopy, imaging, and metrology, among others. In HHG, the highly nonlinear interaction between an intense femtosecond (fs) laser pulse and matter---typically a gas target, though also extended to solids---produces EUV or even soft x-ray harmonic beams \cite{Popmintchev2012}, which are emitted as attosecond pulses \cite{Paul2001,Hentschel2001}. This process is explained by the semiclassial three-step model \cite{Corkum1993, Schafer1993}. First, the laser field ionizes the atomic target. Next, the ejected electronic wavepacket is accelerated in the continuum. Finally, due to the oscillatory behavior of the laser field, the electron recollides with the parent ion. The kinetic energy gained by the electronic wavepacket during the process is released as higher-order harmonics, multiples of the driving frequency. Thanks to the extreme coherence of HHG, the properties of the harmonic beams can be tailored by structuring the fs driving field. For example, the focusing properties of the harmonic beams can be adjusted by driving HHG with spatially shaped laser beams \cite{Quintard2019, Wikmark2019} or by imprinting angular \cite{Hernandez-Garcia2016} or transverse (`lighthouse effect') \cite{Vincenti2012} spatial chirp. Another possibility is to mix two or more driving laser beams, which allows for more degrees of freedom. Indeed, the polarization state of the high-order harmonics was restricted to linear until two driving beams---with proper polarization states and propagation geometries---were used to produce circularly polarized harmonics \cite{Fleischer2014, Hickstein2015} and attosecond pulses \cite{Huang2018}.  

An intriguing approach to control the properties of the high-order harmonics is to drive them with spatially-shaped  profiles rather than a lowest-order Gaussian modes. One of the exciting approaches taken during the last decade has been to drive HHG with Laguerre-Gauss modes, which is the set of free-space modes in the cylindrical coordinate basis. Laguerre-Gauss modes can carry well-defined orbital angular momentum (OAM) that can then be conveyed onto the high-order harmonic beams. Indeed, OAM conservation is seen in the HHG process \cite{Hernandez-Garcia2013, Gariepy2014, Geneaux2016} allowing for the production of harmonic vortices with topological charges as high as one hundred\cite{Pandey2021}. By mixing different Laguerre-Gauss beams carrying distinct OAM, more control can be achieved to produce, for example, circularly polarized high harmonic vortices \cite{Paufler2019, Dorney2019, Pisanty2019}, high harmonic beams with self-torque \cite{Rego2019, delasHeras2024b} or attosecond vortex pulse trains \cite{delasHeras2024a}. 
The other common basis for describing free-space beam solutions is the Hermite-Gauss family of modes (see Appendix A for the full form of these modes). Vector beams, exhibiting a spatially-varying polarization state, can be described as a linear combination of Hermite-Gauss or Laguerre-Gauss modes with different polarization states. Vector beams have also been used to drive HHG in gases \cite{Hernandez-Garcia2017}, plasmas \cite{Venkatesh2022} or thin solids \cite{Garcia-Cabrera2024}. 
Proper analysis of the geometry and topological parameters of such modes, together with the isotropic or anisotropic behavior of the target, allows the study of the preservation of symmetries of the up-converted beam in such a nonlinear process as HHG. For example, whereas in gases the harmonic topological properties mimics that of the driver \cite{Hernandez-Garcia2017}, in graphene the 6-fold symmetry due to its anisotropic behavior is imprinted into the harmonic beams \cite{Garcia-Cabrera2024}, whose topology differs from that of the driver. 

These previous observations inspired us to investigate HHG driven by single-mode Hermite-Gauss beams.  As we will show, due to the non-perturbative nature of HHG, this generation scheme is qualitatively different from the modal mapping in vortex (single Laguerre-Gauss mode) or vector (combination of two Laguerre-Gauss or Hermite-Gauss modes) beam-driven generation. Perturbative harmonic generation driven by Hermite-Gaussian beams was performed in composite thin films (second harmonic generation) \cite{Bernal2004} or in plasmas (third harmonic generation) \cite{Sharma2021} to enhance the harmonic signal. In the non-perturbative regime, two previous works \cite{Camper2014,Camper2015} performed HHG driven by a dual-lobe focal spot resembling an $HG_{0,1}$ Hermite-Gauss beam, produced by inserting a transmissive optic with a 0-$\pi$ phase shift step across the middle of the input beam. This structured beam opened the possibility of providing a two-source HHG interferometry setup, which was used to retrieve the harmonic dipole phase in gases and molecules. 

In this work, we explore experimentally and theoretically the frequency up-conversion of Hermite-Gauss beams in high-order harmonic generation. Our results indicate that, in contrast to the use of Laguerre-Gauss modes or cylindrical vector beams, the Hermite-Gauss mode structure is not directly translated into the generated high-order harmonic beam. Our theoretical and experimental analysis shows that the highly nonlinear process shapes the lobes of the input Hermite-Gauss beam into separate beamlets whose direction is controlled by the local wavefront of the input mode. These beamlets propagate with a divergence strongly influenced by the nonlinear dipole phase and can produce a multi-fringe interference pattern, depending on the relative position between the driving beam focus and the gas jet. 
We carefully characterize this behavior for Hermite-Gaussian modes $HG_{0,1}$ and $HG_{1,1}$. (For $HG_{m,n}$, the subscripts $m$ and $n$ refer to the HG modal indices in the $x$ and $y$ directions, respectively.) To do so, we use a novel passive arrangement in which the driving intense fs mode is derived from a vector beam that is separated into its high-order Hermite-Gauss modes with a polarizer.  Passive, common-path approaches to producing dual \cite{Camper2014,Camper2015} or multiple (this work) high harmonic source beams are in contrast to those in which an input beam is split into multiple paths before being directed to the nonlinear medium (e.g. \cite{Negro2014,Lu2019}). Rather, these experiments fit into an emerging field in which high harmonic pulses must be crossed with attosecond precision and stability to make measurements with interferometric precision. The position of the gas target prior to the focal plane imprints a common, converging wavefront across the Hermite-Gauss driving mode leading to overlap of the high-order harmonic beamlet downstream, in turn resulting in interference either at a secondary target or a camera. We demonstrate a simple application in which we use this self-interference to calibrate the dispersion of an EUV diffraction grating. We also anticipate this method will be used for simultaneously generating multiple, synchronized high harmonic beams to find applications to reducing data acquisition time in scanning ptychography\cite{Goldberger2021} or even to extend single-shot ptychography  \cite{goldberger_three-dimensional_2020, barolak2022} to the EUV.

\section{\label{sec:expt}Experimental measurements of HHG driven with Hermite-Gaussian modes}

\subsection{\label{sec:exptmethods}Experimental methods}

\begin{figure}
{
    \includegraphics[width = 0.475\textwidth]{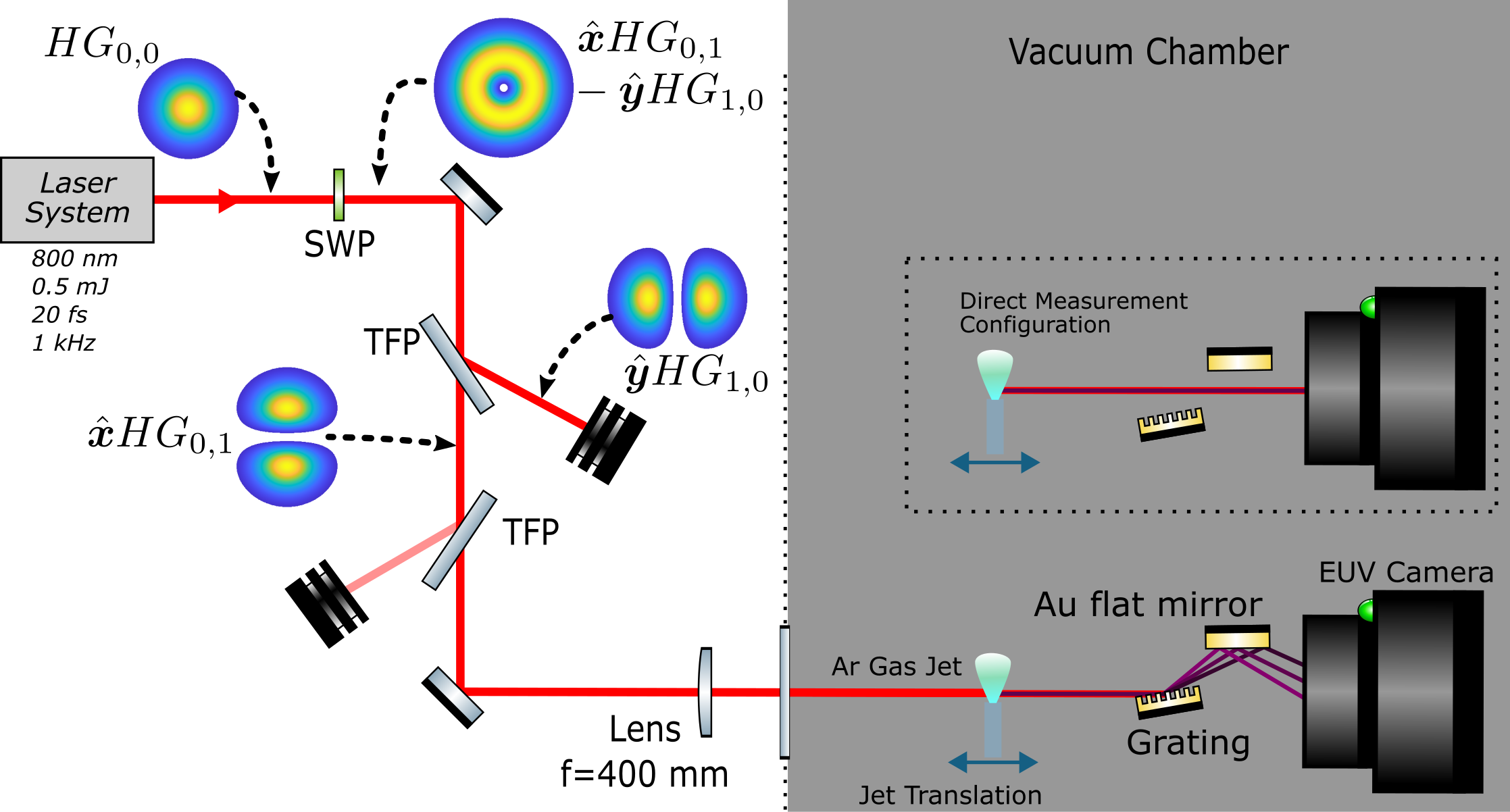} 
  }
  \caption{Experimental setup for HHG driven by Hermite-Gaussian beams. A Gaussian laser beam from the commercial Ti:Sapphire laser, Solstice Ace, is first prepared for the setup through wavefront optimization with a deformable mirror and pointing stability (not shown). The beam is sent through a $p_{vec}=1$ s-waveplate (SWP) to produce an azimuthally-polarized vector beam, which is then polarization-filtered using the thin-film polarizers (TFP) to pass the $HG_{0,1}$ mode. (In later experiments a $p_{vec}=2$ s-waveplate was also used to produce the $HG_{1,1}$ mode). The $HG_{0,1}$ beam is then focused into a vacuum chamber to produce harmonics in a gas jet that can be translated to control the wavefront of the beam in the target. The resulting harmonics are collected directly in the camera or sent into a flat-field grating and directed back to the camera with a gold mirror.}
	\label{fig:HG_VacChamber}
\end{figure}

Figure \ref{fig:HG_VacChamber} is a schematic of our passive scheme for generating infrared (IR) Hermite-Gaussian modes from lowest-order Gaussian beams, in order to drive HHG. We use an S-waveplate (SWP) to convert the linearly polarized Gaussian beam to an azimuthally-polarized vector beam\cite{Beresna2011,Sakakura2020}, where the polarization vector is in the $\pm\hat{\phi}$ direction. As detailed in Appendix \ref{app:appxA}, vector beams of order $p_{vec}$, which have an intensity profile that does not have any modulation in the azimuthal direction, can be polarization filtered to produce beams with well-defined lobes of alternating sign. Using a first-order ($p_{vec}=1$) as shown in Fig. \ref{fig:HG_VacChamber}, we oriented the S-waveplate and thin-film polarizers (TFP) to produce the HG$_{0,1}$ mode for our first experiment. These TFPs (Eksma Optics) have an extinction ratio of 200:1, so that after two elements the polarization purity should be $4x10^4$. In subsequent experiments, we used a $p_{vec}=2$ higher-order SWP to produce a HG$_{1,1}$ mode. Transmission through a sequence of two thin-film polarizers was used rather than reflection to better preserve the beam wavefront, while a pair was used to increase contrast in the filtering. The beam was then focused near the tip of a free-flowing rectangular gas jet mounted on a translation stage to produce the high-order harmonic beams.

\begin{figure*}
  \makebox[\textwidth]{
    \includegraphics[width = 0.85\textwidth]{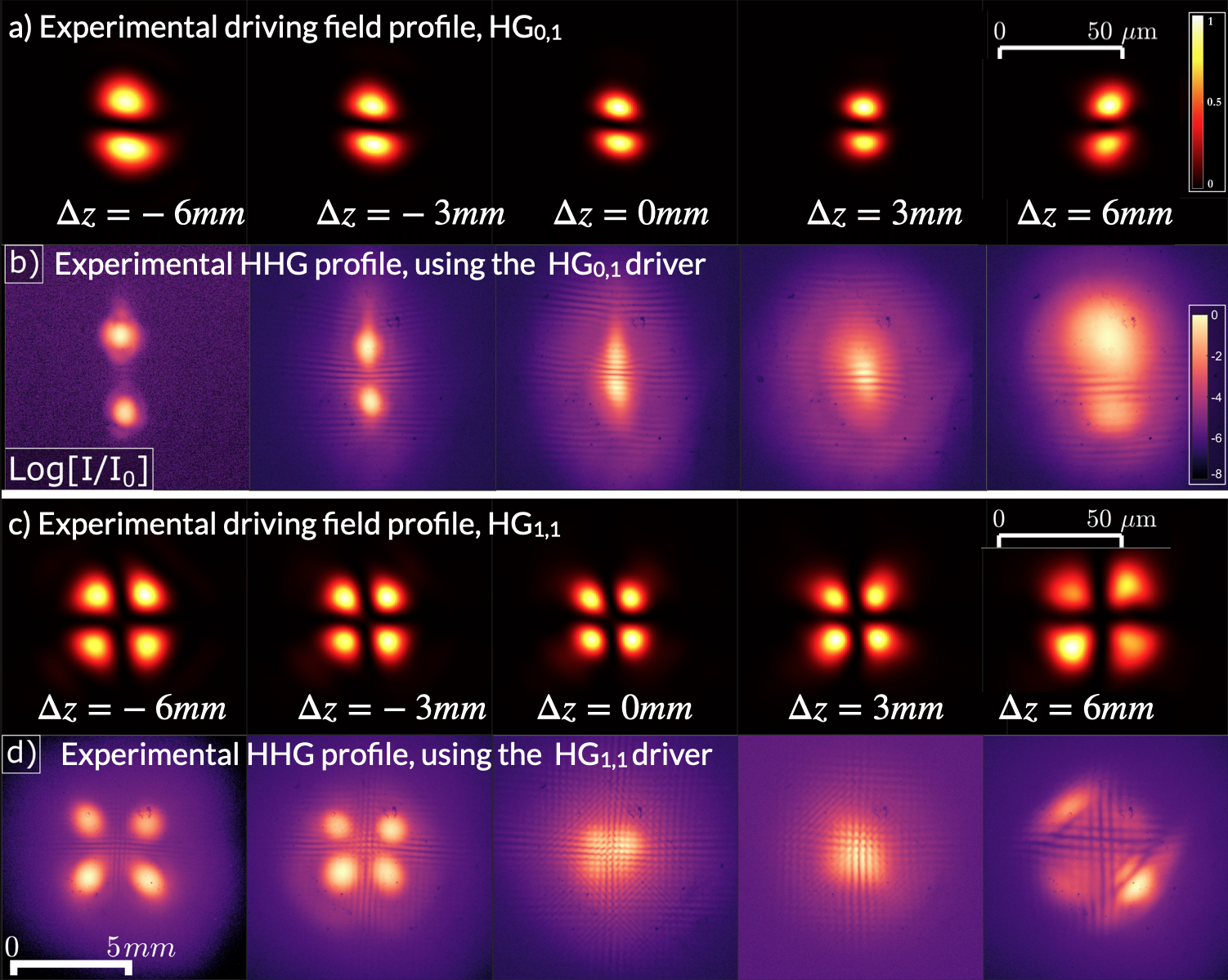} 
  }
  \caption{a,c) Intensity profiles of the fundamental (800 $nm$) $HG_{0,1}$, $HG_{1,1}$ modes, experimentally measured using ptychography for generation at $\Delta z = -6, -3, 0 , 3$ and $6$ mm. b,d) Experimental images of the HHG signal plotted as the logarithm of the normalized signal. The images show the direct HHG beams from generation at five different axial locations denoted by $\Delta z$ measured by an EUV camera placed away from generation. Here, $\Delta z$ is referenced to the approximate location of the beam waist $z = 0$. The focal plane of the 800 $nm$ $HG_{0,1}$ is also not expected to be at the same axial location as the $HG_{1,1}$ focus, since the experimental setup was not preserved between experiments. Camera images taken with a proper exposure time to get a good signal-to-noise ratio. Near the edges this corresponded to 10 seconds of exposure while near the focus it was only several 100 ms.}
	\label{fig:HG_HHGResults}
\end{figure*}

We employed ultrafast pulses from a Spectra-Physics Solstice Ace Ti:sapphire amplifier with a center wavelength of $\lambda=$800 nm, capable of producing up to 6.5 mJ pulse energy. The pulse duration out of the amplifier was measured to be 45 fs using frequency-resolved optical gating. We estimated a duration of 50 fs on target, broadened slightly by third-order phase after grating separation adjustment to compensate for transmissive optics in the system. 

IR pulses from the laser system (typically 0.7 mJ) were directed into the optics to produce the Hermite-Gaussian beam and then focused with a $f$=400 mm lens (Edmund Optics p/n 11664) through a window (1 mm thick) into the vacuum chamber to a Gaussian mode radius of 43 $\mu$m. We used argon gas as the target, irradiated with an approximate intensity of $10^{14} W/cm^2$ in the $HG_{0,1}$ mode. An optional grating-mirror pair can be inserted into the harmonic beam to produce spectrally separated harmonic beams onto the camera that have been partially focused in the horizontal (spectral) direction. The higher-order harmonics produced were gathered directly onto an EUV-sensitive camera after passing through thin Al filters to remove the fundamental beam.

We found that the spatial quality of the high-order harmonic beams driven by high-order spatial modes is very sensitive to wavefront distortion. Therefore, upstream from the SWP we used a deformable mirror and Shack-Hartmann wavefront sensor (Imagine Optic) to flatten the driving beam wavefront. Since the driving beam passed through some additional optics after wavefront correction, small manual corrections to the deformable mirror were performed while monitoring the symmetry of the harmonic beam.

\subsection{\label{sec:resultsHG01}Experimental results of HHG driven with $HG_{0,1}$ and $HG_{1,1}$ modes}
We first discuss the production of high-order harmonic beams from an $HG_{0,1}$ mode. The spatial distribution of the fundamental IR mode is shown in Fig. \ref{fig:HG_HHGResults}a, obtained by back-propagating a ptychography measurement of the fundamental to the focal plane ($\Delta z=0$). The mode profile obtained in this way was consistent with observations of the laser intensity obtained by placing a camera directly at focus. Bypassing the diffraction grating to observe the harmonic beams directly, we varied the position of the gas jet target along the beam direction ($\Delta z$, the nominal distance from the fundamental beam waist or focal plane) through the focal volume with steps of $0.5\ mm$ for a total range of $13\ mm$. The Gaussian mode radius for the focused fundamental was measured to be $w_0=$43 $\mu$m, so this scanning range represents a factor of 0.9 times the confocal parameter ($2 z_R = 14.5 $ mm), where $z_R = \pi w_0^2/\lambda$ is the Rayleigh range. Fig. \ref{fig:HG_HHGResults}b shows a selection of these harmonic beam images. There is no spectral resolution in these images, so the harmonic modes are a superposition dominated by harmonic orders 17-29, as seen in Fig.\ref{fig:HG_SpectralHHG}. The lower end of this range is determined by absorption in neutral argon, while the upper end is set by the harmonic intensity cutoff. For target positions close to the lens, there are two harmonic beamlets that are well defined. Those beamlets move closer together as the jet is moved away from the lens. Where the beams overlap at the camera, close inspection of Fig. \ref{fig:HG_HHGResults}b shows interference between the two harmonics beamlets. For jet positions after the focal plane of the fundamental, the two beamlets are much larger, and interference is still seen on the camera. 

The harmonic beams do inherit some structural features from the fundamental $HG_{01}$ mode: there are still two lobes separated by a node. However, the size of the lobes in the harmonic beams do not fit a single HG mode, rather a large number of HG modes would be required to describe these beams, as will be described in the next section. As noted in the caption, the exposure time was increased when the jet was moved away from the focus. The images are also normalized to the peak harmonic signal so as to maximize the visibility of the harmonic structure. As noted in the caption, the exposure time was increased when the jet was moved away from the focus. The images are also normalized to the peak harmonic signal so as to maximize the visibility of the harmonic structure. 

We extended our approach by replacing the $p_{vec}=1$ s-waveplate in Fig. \ref{fig:HG_VacChamber} with a $p_{vec}=2$ s-waveplate to generate a second-order IR vector beam. This beam was sent through the polarizers to produce a beam very similar to an $HG_{1,1}$ mode (Fig. \ref{fig:HG_HHGResults}c). The expected departures from an ideal mode are described in Appendix \ref{app:appxB}. We used this mode to produce a 2x2 array of high-order harmonic EUV beams (see Fig. \ref{fig:HG_HHGResults}d). The evolution of the output harmonic beamlet spacing and the interference effects follow trends that are similar to the $HG_{0,1}$ scenario shown in Fig. \ref{fig:HG_HHGResults}b: the harmonic beamlets cross and are well-separated in the far field if they are generated prior to the fundamental focus. Then as the jet is scanned through the focus of the fundamental, the harmonics generated begin to produce a high contrast interference pattern, in this case modulating in both spatial directions producing a grid interference pattern. It should be noted that any ellipticity in the initial Gaussian input mode will lead to power imbalance among the lobes of the starting $HG_{1,1}$ mode. The $\Delta z = 6 mm$ position of Fig. \ref{fig:HG_HHGResults}c and d shows a good example of how slight asymmetries in the driving beam lead to large changes in the harmonic profile.

\section{\label{sec:HGmodalcomposition}Understanding the HHG process driven by Hermite-Gaussian modes}

In this section, we present advanced numerical simulations of Hermite-Gaussian-driven HHG, as well as a simple HHG model that allows us to understand the physics behind the experimental results showing self-interference of the harmonic beamlets.

\begin{figure*}
  \makebox[\textwidth]{
    \includegraphics[width = 0.7\textwidth]{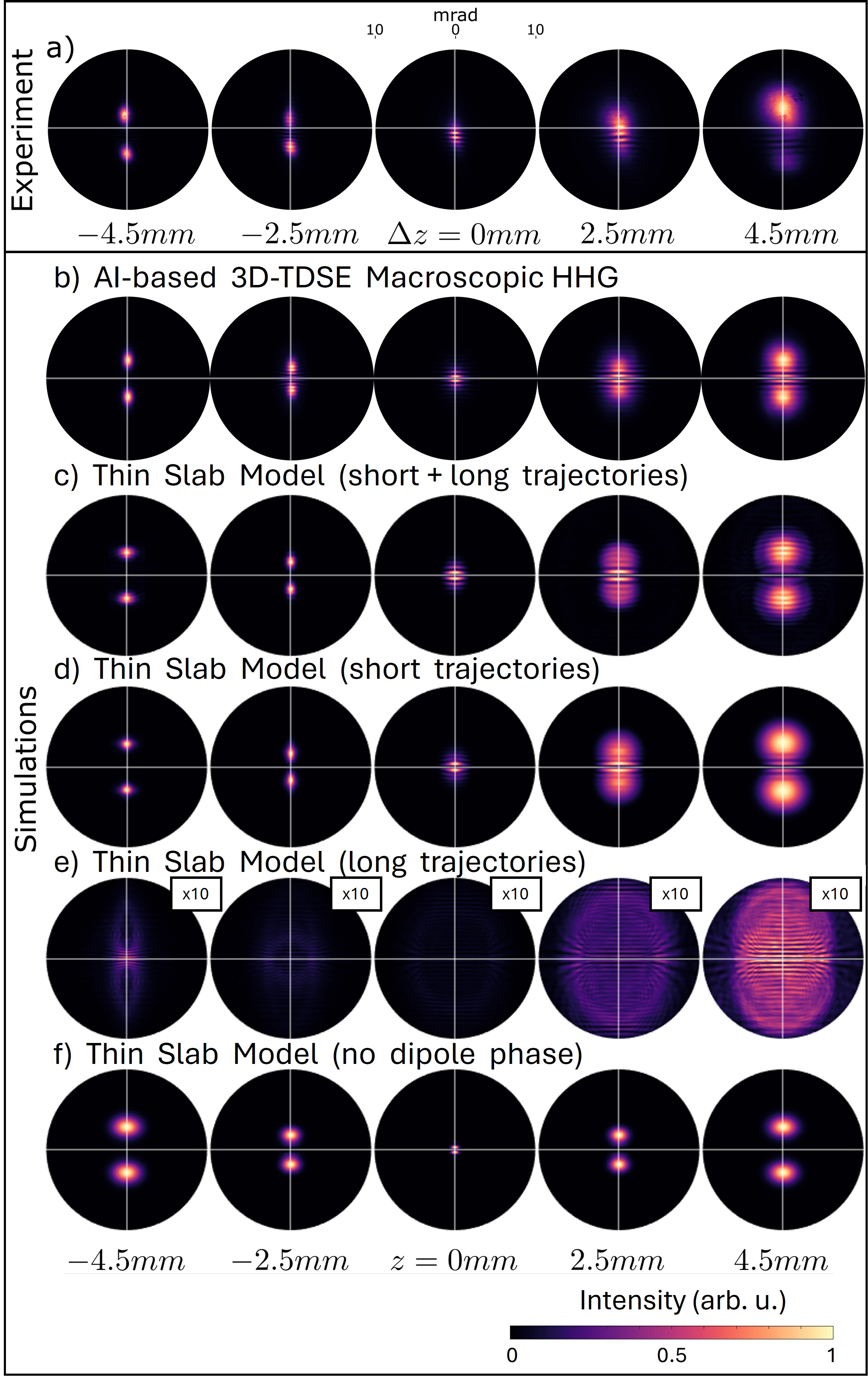} 
  }
  \caption{Theory and experimental comparison of Hermite-Gauss-driven HHG using a $HG_{0,1}$ mode. a) Experimental results of the whole harmonic beam at five positions along the propagation distance. In contrast to Fig. \ref{fig:HG_HHGResults}, the results are plotted in linear scale, along the far-field divergence. $\Delta z = 0$ is not the same center location as in Fig. \ref{fig:HG_HHGResults}, as the experimental data is shifted to match the simulation results. b) Far-field intensity distribution using the numerical AI-based 3D-TDSE macroscopic HHG results in section \ref{AI_model}. c) to f) show the corresponding results using the TSM and taking into account c) short and long, d) only short, and e) only long trajectory contributions within the dipole phase term, f) neglecting the dipole phase term. The long trajectory contributions in e) are artificially enhanced by a factor of 10 to make them visible. The simulations assume an ideal $HG_{0,1}$ with waist parameters matching the experiment. $z$ for the simulation results are absolute and so the center harmonic image was generated at the focus of the ideal $HG_{0,1}$ beam.}
	\label{fig:HG_HHG_Sim}
\end{figure*}

\subsection{Numerical modeling of Hermite-Gaussian-driven HHG via an artificial-intelligence-based 3D-TDSE method}
\label{AI_model}

The theoretical description of Hermite-Gaussian-driven HHG requires a comprehensive description of both the microscopic---laser-driven quantum wave-packet evolution---and macroscopic---harmonic phase-matching---perspectives. To achieve this, we used an artificial intelligence (AI)-based method in which the dipole acceleration is predicted by a neural network trained on the three-dimensional time-dependent Schr\"odinger equation (3D-TDSE), while macroscopic phase-matching is considered through the integral solution of the Maxwell equations \cite{PablosJ2023}. Input data for the simulation, such as the spatial amplitude and phase profiles of the driving field, were inferred from experimental measurements. We performed a ptychographic characterization of the 800 $nm$ fundamental beam within our setup, as described in Goldberger et al\cite{Goldberger2021}. This technique provided the amplitude and wavefront profiles at the measurement plane, located approximately 40.16 cm from the diagnostic camera. These profiles were used to numerically reconstruct the beam properties at various positions throughout the focus. This information verified the Hermite-Gaussian modal structure and determined the beam parameters used in the AI-based 3D-TDSE simulations. The beam waist was measured to be $w_0 = 43\, \mu m$. The numerical back-propagation of the fundamental beam to $\Delta z = 0 $ of the measured fundamental beam is illustrated in Figs. \ref{fig:HG_HHGResults}a and \ref{fig:HG_HHGResults}c for the $HG_{0,1}$ and $HG_{1,1}$ modes respectively. 

The neural network that predicts the dipole acceleration at each position within the gas jet, was trained on a dataset comprising $8\times 10^4$ 3D-TDSE calculations performed in Ar. The driving pulse duration was modeled with a $\sin^2$ envelope, a central wavelength of 800 $nm$ and pulse duration of $7.7$ fs in FWHM in intensity. The range of spatial phase and peak intensities considered in the dataset varied from 0 to $2\pi$, and from $3.81 \times 10^{13}$ W/cm$^2$ to $1.97 \times 10^{14}$ W/cm$^2$, respectively. We note that the modeled pulse duration was shorter than that used in the experiments, due to computational time constrains. However, this discrepancy is not expected to introduce fundamental deviations in our results. 

The neural network was trained on the spectral domain of the HHG signal, separated in real and imaginary parts. The training dataset 3D-TDSE calculations were performed with a temporal grid comprising 8192 points and a grid spacing of 0.94 attoseconds, and a spatial grid in cylindrical coordinates with 3200 points in the direction of the field polarization and 400 points in the radial direction, with a resolution of $5.3 \times 10^{-3}$ nm. With those parameters, the noise level in the normalized HHG spectrum was approximately $10^{-5}$ with respect to the harmonics in the plateau region of the spectrum. Note that to optimize the training of the high-order harmonics measured in the experiment, the HHG spectra were filtered below the 12th harmonic order.

The training and validation of the neural network were conducted using Keras and TensorFlow libraries\cite{chollet2021deep}, using Adam as the optimizer and the Mean Squared Error (MSE) as the loss function. A progressive batch size strategy was implemented to improve generalization and stability during training. The final trained model achieved an error metric (MSE) on the order of $10^{-5.5}$ when compared to direct 3D-TDSE calculations. More details about neural network's architecture, training, validation and computational time required to perform a full macroscopic calculation can be found in \cite{PablosJ2023}.

Fig. \ref{fig:HG_HHG_Sim} shows a direct comparison between the (a) experimental and (b) numerical results using the AI-based 3D-TDSE model, showing very good agreement on the main features. The model correctly shows the generation of two localized harmonic beamlets that approach each other at the camera as the jet is moved from close to the lens. The beamlets cross and interfere, then ultimately diverge at large, positive $\Delta z$. The asymmetric behavior of the harmonic beamlet profiles when changing the gas jet position with respect to the focus is theoretically reproduced---i.e. when comparing results obtained at $\pm\Delta z$. The comparison when using the $HG_{1,1}$ driving mode is shown in Fig. \ref{fig:HG11_HHG_Sim} in Appendix \ref{app:appxHG11}, showing similar qualitative agreement between the model and experiment. There are some discrepancies, especially at large $\Delta z$, where the experimental beams are more asymmetric than the model. We attribute this to the enhancement of slight imperfections in the driving beam symmetry by the highly nonlinear process. In the next subsections we analyze, through simpler physical models, the origin of the features observed both in the experiment and in the numerical simulations. 

\subsection{\label{sec:TSMmodel}Simple perturbative and non-perturbative models of Hermite-Gauss driven HHG. The Thin Slab Model}

Unlike in Laguerre-Gauss- or vector beams-driven HHG \cite{Hernandez-Garcia2013, Hernandez-Garcia2017}, for the Hermite-Gaussian beam input, it is not as clear what symmetries may be preserved in the HHG process, since there is not an azimuthal symmetry that is smooth in amplitude. Understanding how symmetries in the driving beam are or are not preserved during HHG may be understood by using a simple, Thin-Slab Model (TSM) of the process \cite{Hernández-García_2015, Rego2016}. Here, we will show how the non-perturbative behavior affects Hermite-Gauss-driven HHG for non-depleted pump conversion in which phase-matching is not a major factor. The TSM uses a reduced description for the $q^{th}$-order harmonic generation in which the yield depends on the $q_{eff}^{th}$-power of the fundamental amplitude, and there is an intensity-dependent dipole phase, $\phi_{dp}(I_{fund})$. This dipole phase is related to the time taken for the electron wavepacket to return to the ion following tunneling ionization, therefore it is different for short and long trajectory contributions \cite{Hernández-García_2015}. We can express the generated $q^{th}$-order harmonic field, $E_{q}$, as: 
\begin{equation} \label{Equ:HHGEfield}
E_{q}(x,y,z_G) \propto |E_{fund}(x,y,z_G)|^{q_{eff}} e^{-iq\phi_{fund}(x,y,z_G)-i\phi_{dp}(x,y,z_G)},
\end{equation}
where $E_{fund}$ is the electric field of the fundamental driving laser, $\phi_{fund}$ is its phase, and $z_G$ is the axial location of the generation process. For the perturbative case, $q_{eff}=q$ and $\phi_{dp}=0$. For non-perturbative HHG, the parameter $q_{eff}$ is found to be around four for the higher-order harmonics, though it slightly varies from harmonic to harmonic \cite{Durfee1999,Rego2016,Dacasa2019}. 
Since the dipole phase depends on the intensity structure of the fundamental, the wavefront shape of the harmonic results from the sum of $q\phi_{fund}(x,y,z_G)$ and the intensity-dependent phase, $\phi_{dipole}(x,y,z_G)$. 

While an analytic solution is achievable for certain scenarios, we can use Eq. \ref{Equ:HHGEfield} and use numerical propagation to look at the properties of the generated harmonics. In the TSM, we assume for simplicity that the signal is generated in a thin slab, which neglects longitudinal phase-matching effects \cite{Hernández-García_2015}. We can use Eq. \ref{Equ:HHGEfield} as the source and use Fraunhofer propagation to calculate the far-field output for a given harmonic order. 
Results for the TSM when HHG is driven by a $HG_{1,0}$ mode are shown in Figs. \ref{fig:HG_HHG_Sim}c to \ref{fig:HG_HHG_Sim} f, for different values of the dipole phase. These results include the superposition of harmonic orders 15-31, whose weights are chosen according to the experimental measurement. Fig. \ref{fig:HG_HHG_Sim}c considers both short and long trajectory contributions, showing excellent agreement against the TDSE-based numerical results (Fig. \ref{fig:HG_HHG_Sim}b) and the experimental data (Fig. \ref{fig:HG_HHG_Sim}a). Small deviations in divergence values between TDSE-based and TSM results are attributed to the choice of $q_{eff}$, which, for simplicity, was set to four for all harmonic orders.To disentangle the role of short and long trajectory contributions, we present TSM results including only short (Fig. \ref{fig:HG_HHG_Sim}d) and only long Fig. (\ref{fig:HG_HHG_Sim}e, enhanced by a factor of 10 for visibility) trajectory contributions. Additionally Fig. \ref{fig:HG_HHG_Sim}f shows the case where both are neglected, i.e. we deliberately omitting the dipole phase term. The results clearly indicate that the short trajectory contributions dominate, while long trajectory contributions, being weaker, primarily introduce subtle horizontal fringes at $\Delta z >0$. Furthermore, the dipole phase term is strongly responsible for the asymmetric behavior of the harmonic profiles when moving the gas jet with respect to the focus position, i.e. when comparing results obtained at $\pm\Delta z$. Indeed, the self-interfering profiles obtained at $\Delta z >0$ highlight the strong influence of the dipole phase. These conclusions are also obtained when driving HHG with a $HG_{1,1}$ mode, as shown in Fig. \ref{fig:HG11_HHG_Sim}.

\subsection{\label{sec:beamletDiscussion}High-harmonic beamlet propagation}

\begin{figure}
{\includegraphics[width = \columnwidth]{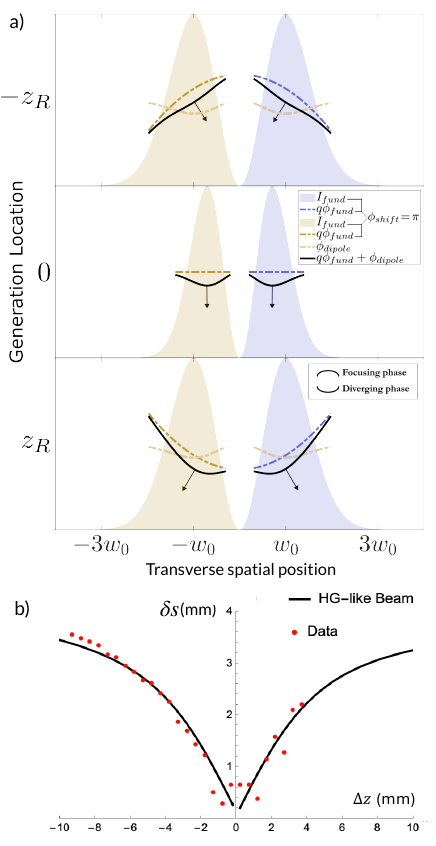}}
\caption{a) Schematic illustration of Hermite-Gaussian driven HHG at three different axial generation locations. The plot shows the intensity of the fundamental driving Hermite-Gaussian mode as a filled-in area along with a phase line for the fundamental beam. The beam area and phase line are plotted in similar colors for different sides of the Hermite-Gaussian mode to denote a relative phase offset of $\pi$ that has been removed in these plots to show symmetry. The intensity-dependent dipole phase is plotted on each beam. The solid black line represents the combined total phase of the harmonic. An arrow is plotted to show the local wavefront direction of propagation after generation.  b) Harmonic beamlet separation ($\delta s$) at the camera (red dots) as a function of the jet position $\Delta z$ relative to the focal plane. The solid line represents a separation calculated from the local slope of the fundamental wavefront.}
\label{fig:HG_1DCartoon}
\end{figure}

The evolution of the high-order harmonic beams presented in the experimental and theoretical results shows that while the harmonic beams inherit the nodal lines that separate the beam into lobes, the actual field distributions are not pure Hermite-Gauss modes. A single Hermite-Gauss mode will propagate self-similarly, with its field distribution simply scaled by the evolving mode size $w(z)$. Instead, we see that the beam propagation evolves in a way that supports the view that the HHG process is producing two phase-locked beamlets of harmonics. Figure \ref{fig:HG_1DCartoon}a illustrates a simple view of this process for the $HG_{0,1}$-driven harmonics. The intensity of the driving beam ($I_{fund}$)is plotted as a filled-in profile for the driving beam and the lines schematically represent wavefronts. The difference in colors represents the $\pi$ phase shift between the two beamlets. For a gas jet position at the fundamental focus ($z = 0$, center row), the wavefront of the fundamental ($\phi_{fund}$, colored dash-dot lines) is flat, and the wavefront of the harmonics (solid) is dominated by the nonlinear dipole phase ($\phi_{dipole}$, colored dashed lines). For a gas jet position closer to the focusing lens, the converging, common wavefront of the fundamental can balance the diverging contribution of the nonlinear dipole phase, leading to smaller harmonic beamlets in the far field. These beamlets are directed with an angle that corresponds to the local slope of the common wavefront, evaluated at the centroids of the beamlets. For gas jet positions that are after the focus, the two contributions to the wavefront are both diverging. 

This view is supported by Fig. \ref{fig:HG_1DCartoon}b, where we compare the measured spot separation at the camera for the input $HG_{0,1}$ mode (red dots) with what is predicted by calculating the local slope of the fundamental wavefront (solid black line). In regions where there is interference between the beamlets on the camera, the beamlet centroids were calculated by using the Fourier transform. The geometric crossing plane will be at the center of curvature of the input wavefront, $z_{cr} = z_R^2/z_{jet}$, where $z_{jet}>0$ is the jet distance in front of the beam waist. Using the peak location instead of the centroid of the beamlet, the crossing angle is then ArcTan$[\sqrt{2} w_0 (z_{jet}/z_{cr}]$ for the $HG_{1,0}$ mode. For the $HG_{1,1}$, the $\sqrt{2}$ factor is absent since the diagonal distance is used for the transverse starting location from the axis. The calculated crossing angles that fit the data best corresponded to a $w_0=36$ $\mu m$, which is about 16\% smaller than measured. This discrepancy is consistent with the slight departure of the focused beam from a perfect $HG_{0,1}$ mode, as described in Appendix \ref{app:appxB}. The good agreement of our measurements with the beamlet directions calculated from the wavefront confirms our simple picture of the process, where the dipole phase strongly affects HHG driven by Hermite-Gauss beams.

\begin{figure*}
  \makebox[\textwidth]{
    \includegraphics[width = 0.6\textwidth]{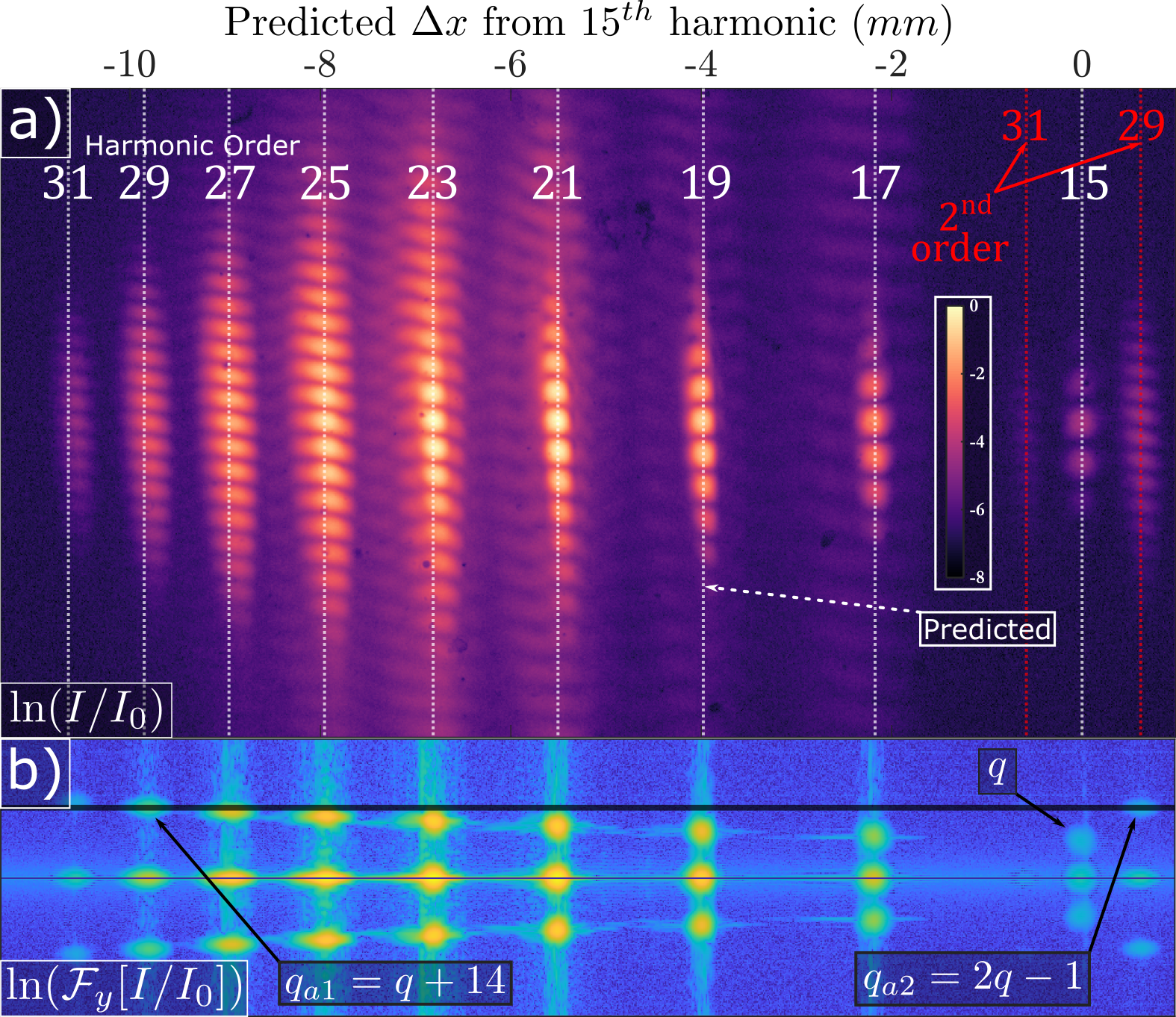} 
  }
  \caption{a) Log of the spectrally-resolved HHG image experimentally collected from the detector. (b) Log of the 1D Fourier transform in the vertical (spatial) direction of the data in (a). Harmonic order $q$ is seen to be next to a $2^{nd}$-order peak $q_{a2}$ that has the same fringe density as harmonic $q_{a1}$. White dotted lines are plotted in (a) for the predicted spatial location, $\Delta x$, of all the other harmonic orders based on the grating angle and distance between the grating and the detector.}
	\label{fig:HG_SpectralHHG}
\end{figure*}

\section{\label{sec:spectcal}Calibrating the spectral dispersion of a EUV diffraction grating through Hermite-Gauss-driven HHG}
In Fig. \ref{fig:HG_HHGResults}, we observed interference where the two high-harmonic beamlets are overlapping at the camera. Similar "self-interference" has been observed in earlier experiments where a $\pi$ phase-step mask was used to generate an $HG_{1,0}$-like beam at the focus and has been applied to interferometric measurements of gases and molecules \cite{Camper2014, Camper2015}. Here, we apply this interference to address a challenge common to experimentalists who use a free-standing grating to disperse the high-order harmonics, namely that it is difficult to know how to assign harmonic order numbers to the peaks of the harmonic comb. The free-space mounting approach uses the small source point in the jet target instead of an entrance slit, and the grating-camera distance  and incident and diffracted angles are not trivial to measure to an accuracy sufficient to unambiguously associate the harmonics orders to the detected peaks.  While it is clear that the energy spacing of the harmonics is twice the fundamental photon energy, the spacing at the camera plane varies weakly with harmonic order spacing.

Based on our understanding that the direction of the harmonic beamlets is controlled by the local slope of the fundamental wavefront, all harmonic orders should be launched with the same initial angle. We show here that the self-interference effect can be used to identify the harmonic orders, taking advantage of the fact that at a given jet position the source point separation is independent of harmonic order. We are using a flat field grating with a variable line spacing to image the spectrum to a flat field (McPherson 251MX, 300 grooves/mm). A flat Au-coated mirror directs the spectrum to the camera position that is used when the harmonics are observed directly (Fig. \ref{fig:HG_VacChamber}). We gathered spectrally separated images for the same generation points scanned for the data shown in Fig. \ref{fig:HG_HHGResults}. An example of one of these images corresponding to $\Delta z = 2$ $mm$ can be seen in Fig. \ref{fig:HG_SpectralHHG}.a. In the spectrum, we can see signals from the $1^{st}$ and $2^{nd}$ diffracted orders. Near one of the low-order harmonics ($1^{st}$ grating order), we can see two neighboring low-power signals. By doing a 1D Fourier transform, seen in Fig. \ref{fig:HG_SpectralHHG}b we can see that there is a $1^{st}$-order diffraction peak of a low-order harmonic (marked $q$ on the figure) that is next to a $2^{nd}$ order diffracted peak of a higher harmonic that we will call $q_{a2}$. By looking at the fringe densities in Fig. \ref{fig:HG_SpectralHHG}b, we can identify which $1^{st}$-order harmonic $q_{a1}$ has the same wavelength as $q_{a2}$, finding that $q_{a1} = q+14$. Since $q_{a2}$ is at a lower-wavelength position on the camera than $q$, we can say that $q_{a2} = 2q-1$. Since $q_{a1}=q_{a2}$ we can unambiguously identify that $q = 15$ in this case, allowing the rest of the harmonic orders to be labeled in the image. As a secondary check, we can predict the spatial separation on the camera ($\Delta x$) of the harmonics relative to the $15^{th}$ harmonic based on the grating equation and the distance between the grating and the camera. This prediction lines up with the image and provides further confidence in our spectral calibration.

\section{\label{sec:ptych}Single-Shot Ptychographic Imaging using Hermite-Gaussian HHG Beams}

\begin{figure*}
   \makebox[\textwidth]{
     \includegraphics[width = 0.75\textwidth]{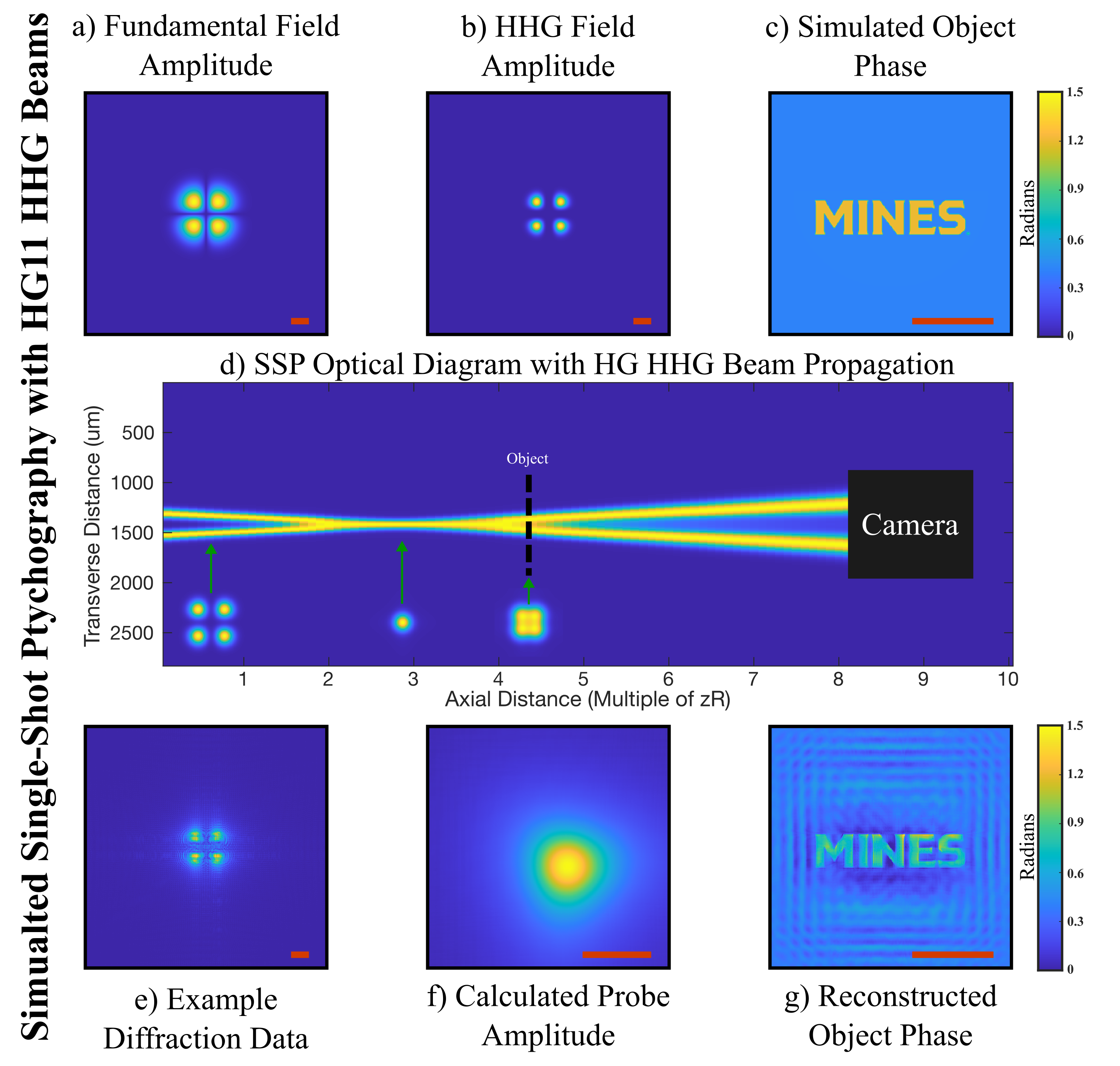}
   }
   \caption{Simulation results and optical schematic of single-shot ptychography (SSP) using Hermite-Gaussian HHG light. In this proposed microscope design, the Hermite-Gaussian HHG beam acts as the probing beamlet array for SSP. The optical schematic shows the simulated propagation of an $HG_{1,1}$ HHG beam starting from the generation plane and ending at the detection plane. Underneath the propagation, the beam profile is shown at the generation plane, the cross-over plane, and the object plane. In the optical diagram, the interference fringes have been blurred for plotting purposes to more easily see the overlap of the beamlets as a function of axial distance. In the upper left and middle, the fundamental and HHG field amplitudes are shown. c) shows the simualted object phase profile which is a thin sheet of aluminum in the shape of letters spelling "MINES". e) shows the diffraction data at the camera plane, plotted to the half power. f) shows the probe beamlet amplitude at the object plane for one probing position. g) shows the reconstructed object phase profile, displaying the MINES letters. The red scale bar in each image is 180 $\mu m$.}
 	\label{fig:SSP}
\end{figure*}

Ptychography is a powerful coherent diffractive imaging technique that simultaneously reconstructs the amplitude and phase profile of both an object and the probing illumination from a set of diffraction intensity patterns probed at overlapped locations of the object. Owing to its lensless optical setup, ptychography is a particularly useful imaging modality for spectral ranges where aberration-free imaging optics are challenging to produce. Ptychography using HHG light has recently been of interest to the scientific community, having been shown to image very small features, highly periodic structures with high fidelity, and over a wide spectral range \cite{seaberg_ultrahigh_2011,loetgering_advances_2022,tadesse_wavelength-scale_2019,eschen_structured_2024}. Ptychography has also been shown to be a powerful beam metrology technique for HHG light, where instead of focusing on the object reconstruction, ptychography is used directly to measure the spatial amplitude and wavefront profile of HHG beams \cite{thibault_probe_2009,goldberger_spatiospectral_2021,kewish_reconstruction_2010,david_dwight_schmidt_exploring_2023}. However, the scanning requirement in ptychography leads to long data acquistion times, which can result in lower fidelity results from beam instability, and precludes the ability to image dynamic, difficult to repeat phenomena. Single-shot ptychography (SSP) has been shown, in optical regimes, to image simultaneously with multiple wavelengths, polarization states, in three spatial dimensions, and even as a way to generate ultrafast motion pictures of single-event conduction-band electron dynamics \cite{barolak2022,goldberger_single-pulse_2022,goldberger_three-dimensional_2020,chen_multiplexed_2018,wengrowicz_experimental_2019,barolak_ultrafast_2024, panSingleShotPtychographical2013,sidorenkoSingleshotPtychography2016,kharitonovSingleshotPtychographySoft2022}. With shorter probing wavelengths using HHG light, the ability to extend these SSP techniques to nanoscopic dynamics becomes possible. Here we propose, to the best of our knowledge, the first single-shot ptychography microscope for HHG light.

In standard optical SSP techniques, multiple beamlets are generated using a diffractive optical element and focusing optics are used to generate spatially separated diffraction patterns, from overlapped probing regions of a sample, on a 2D pixelated detector. To overcome the need for a diffractive optical element and focusing optics, our novel HHG SSP method utilizes the Hermite-Gaussian HHG light beams to generate a probe array of individual beamlets that spatially overlap at one axial location and separate at another axial location. In this design, we focused on using an $HG_{1,1}$ fundamental beam to generate an HHG beam with four individual lobes which act as the beamlets for our ptychographic microscope, however a different order Hermite-Gaussian fundamental beam could be used. To tailor the propagation profile of the HHG beamlet array and the focus of each individual beamlet, the parameters of the fundamental beam can be varied. The local wavefront at the generation plane will determine the crossing angle of the HHG beamlets, therefore by changing the distance away from focus of the fundamental beam one can control the crossing angle of the HHG beamlet array. This is necessary as the fundamental beam can focus at a different axial location than the plane of the object which can prevent damage to the object due to the high intensity of the fundamental beam at focus. The width of the individual beamlets at both the cross-over plane as well as the far field depends on both the size of the fundamental beam waist and, due to the dipole phase term, the intensity of the fundamental light at the generation plane. By placing the object some distance outside the cross-over plane, the desired level of overlap of the beamlets can be achieved \cite{huangOptimizationOverlapUniformness2014}. The far field diffraction patterns can then be collected a known distance after the object, once the beamlets have spatially seperated. An optical schematic of the microscope and HHG propagation is shown in Fig. \ref{fig:SSP}. In the camera plane, the diffraction patterns are collected simultaneously in a single camera image and algorithmically chopped out to form the ptychographic dataset. The reconstruction is then performed assuming there is no coherent cross-talk between adjacent diffraction data. However, a more sophisticated algorithm could be developed to account for this coherent cross-talk. The reconstruction resolution is generally given by $dx_{obj} = \lambda_{fund}\;z / (q\;dx_{det}\;N_{det})$ where $dx_{obj}$ is the reconstructed pixel size, $q$ is the harmonic number, $z$ is the object-detector distance, and $dx_{det}$ and $N_{det}$ are the detector pixel size and number of pixels in a single dimension of the detector with diffraction data above the noise level.

We developed a simulation of our SSP microscope using HHG light to verify the capability of these techniques. The simulated fundamental beam had an $HG_{1,1}$ beam profile with a 100 $\mu$m beam waist and an 800 nm central wavelength. The 15\textsuperscript{th} harmonic  was produced 3.9 $cm$ away from the focus with a pulse energy of 425 $\mu$J and a Gaussian temporal profile with a 40 fs pulse duration. The HHG light was generated using equation \ref{Equ:HHGEfield}. Figure \ref{fig:SSP} shows the simulated fundamental beam and $HG_{1,1}$ HHG beamlet array intensity profiles. The simulated object was a 10 nm thick sheet of aluminum in the shape of the word "MINES", with the inside of the letters being aluminum and the outside being free space. The simulated object was placed 3.5 $z_R$ after the generation plane. The example diffraction data collected in the far field is shown in figure \ref{fig:SSP}. The simulated camera had 1000 by 1000 pixels with 2 $\mu$m square pixels. The data was reconstructed using an ePIE algorithm with super-resolution capabilities to help account for the fact that the spatial frequency spectrum is unevenly sampled by having non-centered diffraction data \cite{maiden_superresolution_2011-1}. The reconstructed object phase is shown in figure \ref{fig:SSP} and clearly shows the "MINES" lettering that was simulated. In this simulation, the probes were calculated and supplied to the reconstruction algorithm as a constraint. This can be experimentally achieved by pre-characterizing the probing illuminations as was done by Chang \textit{et al} \cite{chang_single-shot_2020}. For cases where the system does not produce harmonics with sufficient shot-to-shot stability, probe recovery can be included in ways that have been previously demonstrated\cite{levitan2020,Levitan2025}. More details on the simulation parameters can be found in the thesis by Barolak \cite{barolak_thesis_2024}. This novel SSP technique has the potential to bring nanoscopic single-shot imaging capabilities to table-top EUV sources.

\section{\label{sec:conclusion}Conclusions}
Our experimental and theoretical results demonstrate the richness of Hermite-Gaussian-driven HHG, where the properties of the generated harmonics depend strongly on the interaction configuration. Through the development of simple theoretical models, we have shown the relevant role of the dipole harmonic phase in the generated harmonic profiles. In particular, when the gas jet is placed near and after the focal plane, harmonic beamlets are seen to self-interfere at the camera, in the sense that the harmonic beam created from a single, well-defined HG fundamental mode has a structure that allows different portions of the harmonic beam to interfere at a downstream position. Thus this approach allows for the production of an array of harmonic beamlets that is particularly useful for applications. In particular, Hermite-Gaussian-driven HHG presents unique opportunities to explore high contrast and stable interferometry in the EUV spectral range. We have shown that increasing the vector beam order before the polarizer makes it possible to produce four neighboring beams, thereby obtaining a 2D interference grid. The Hermite-Gaussian-driven HHG method also provides an interesting possibility for EUV transient grating experiments and single-shot ptychography experiments. HHG beams have been used in transient grating experiments as the probe but they are generally too low power to be used as the grating \cite{Foglia2023}. More powerful free-electron laser sources of EUV light have been used for investigating transient phenomena with transient gratings\cite{Mincigrucci2018}. While it will be important in the future to study how phase matching can optimize the driving intensity and target pressure to maximize the harmonic flux, it is an exciting prospect to push the limits of HHG to reach the required intensities to perform EUV imaging with our method that does not require a free-electron laser facility.  Hermite-Gaussian-driven HHG represents a possible avenue for generating these gratings within a high-flux HHG system since the input beam parameters can be made into a well-defined mode.

\begin{acknowledgments}
The Colorado School of Mines authors gratefully acknowledge funding from the Air Force Office of Scientific Research through grant no. FA9550-22-1-0495. Universidad de Salamanca authors acknowledge funding from the European Research Council (ERC) under the European Union’s Horizon 2020 research and innovation programme (grant agreement No 851201), from grant PID2022-142340NB-I00 financed by Ministerio de Ciencia e Innovacci\'on and Agencia Estatal de Investigaci\'on, from Junta de Castilla y Le\'on and Fondo Europeo de Desarrollo Regional (FEDER), under grant No. SA108P24, and RES resources provided by BSC in MareNostrum 5 and CESGA in Finisterrae III to FI-2024-2-0010 and FI-2024-3-0035. The University of Southampton authors acknowledge support from the European Research Council (ENIGMA, 789116). 
\end{acknowledgments}

\section{Disclosures}
The authors declare no conflicts of interest.

\section*{Data Availability Statement}
Data underlying the results presented in this paper are not publicly available at this time but may be obtained from the authors upon reasonable request.

\appendix

\section{Hermite-Gauss projection analysis}
\label{app:appxA}

In this appendix section, we describe various aspects of modal decomposition onto Hermite-Gauss modes: conversion of vector beams to Hermite-Gauss beams, evaluation of the modal purity for different approaches to generate particular Hermite-Gauss modes, and to understand how the high-order nonlinear polarization projects onto Hermite-Gauss modes. 

A 2-dimensional Hermite-Gaussian beam can be written as 
\begin{eqnarray}
    HG_{l,m}(x,y,z) = E_{0}^{HG} H_l\left(\frac{\sqrt{2} x}{w(z)}\right) H_m\left(\frac{\sqrt{2} y}{w(z)}\right)\nonumber\\ G(r) e^{\displaystyle i \psi_{lm}(z) }    
    \label{Equ:HGmode}
\end{eqnarray}
where $(l,m)$ are the $(x,y)$ mode indices, $r^2=x^2+y^2$ and $H_i$ is the $i^{th}$ Hermite polynomial. These modes share a common fundamental Gaussian envelope 
\begin{equation}
    G(r) = \frac{w_0}{w(z)} e^{\displaystyle-\frac{r^2}{w^2(z)}} e^{i\left(k z + \frac{k r^2}{2 R(z)} \right)}, 
\end{equation}
where we use the $exp[\displaystyle-i\omega t]$ sign convention. The radii of the mode amplitude and wavefront are defined as $w(z)=w_0 \sqrt{1+z^2/z_R^2}$ and $R(z)=z(1+z_R^2/z^2)$, where $z_R$ is the Rayleigh length, $k$ and $\lambda$ are the wavenumber and wavelength. 

For the Hermite-Gaussian beam, the Gouy phase is defined as
\begin{equation}
    \psi_{lm}(z)= \displaystyle-(l+m+1)\arctan\left(\frac{z}{z_R}\right).
\end{equation}

Hermite-Gaussian beams have a polarization state that is spatially uniform. Vector beams have a spatially-dependent polarization state. Simple examples of vector beams are radial- and azimuthally-polarized beams, where the polarization is locally linear, but has an orientation that is proportional to the azimuthal angle $\phi$. The Jones vector representation of a radially-polarized polarization state is $\{\cos [\phi], \sin [\phi]\}$ while for an azimuthally-polarized state is $\{-\sin [\phi], \cos [\phi]\}$. In our approach to modal conversion from Gaussian to a particular higher-order Hermite-Gaussian mode, we pass a vector beam through a polarizer. For example, a radially-polarized vector beam can be written as the following linear combination of HG modes: $\mathrm{HG}_{1,0}(x, y) \hat{\boldsymbol{x}}+\mathrm{HG}_{0,1}(x, y) \hat{\boldsymbol{y}}$. Similarly, an azimuthally-polarized beam would be written as $-\mathrm{HG}_{1,0}(x, y) \hat{\boldsymbol{y}}+\mathrm{HG}_{0,1}(x, y) \hat{\boldsymbol{x}}$. Therefore, passing a vector beam defined this way through a linear polarizer will select an $HG_{1,0}$ mode that is rotated with respect to the orientation of the polarizer with 50\% efficiency. 

In general, a high-order vector beam is produced by rotating linear input polarization by a factor $p_{vec}$ times the azimuthal angle $\theta$. The Jones matrix representation of this operation is a rotation matrix that rotates the input vector by $p_{vec} \theta$:

\begin{equation}
    \left(\begin{array}{cc}
\cos p_{vec} \phi & -\sin p_{vec} \phi \\
\sin p_{vec} \phi & \cos p_{vec} \phi
\end{array}\right). 
\end{equation}
While spatial light modulators can be used for mode conversion, an alternate approach is to take advantage of waveplates that convert a linearly polarized beam to a vector beam (e.g. radially or azimuthally polarized). Such a waveplate that is particularly useful with high damage threshold and over a wide wavelength range can be produced by micromachining birefringent features inside bulk fused silica\cite{Beresna2011,Sakakura2020}. A waveplate that locally rotates the polarization to convert a linearly polarized input to a vector beam is sometimes called an S-waveplate (SWP)\cite{Beresna2011}. Rotation of the SWP smoothly changes the vector polarization from radial to azimuthal. The SWP is a flexible platform for mode conversion, since vector and vortex beams can be written as linear combinations of Hermite-Gaussian beams\cite{Maurer2007,Kimel1993,Tidwell1993}. Since the SWP is composed of microscopic half-wave retarders, the local half-wave retarder rotation angle is $\theta = p_{vec} \phi/2$.

\begin{figure}
\includegraphics[width = \columnwidth]{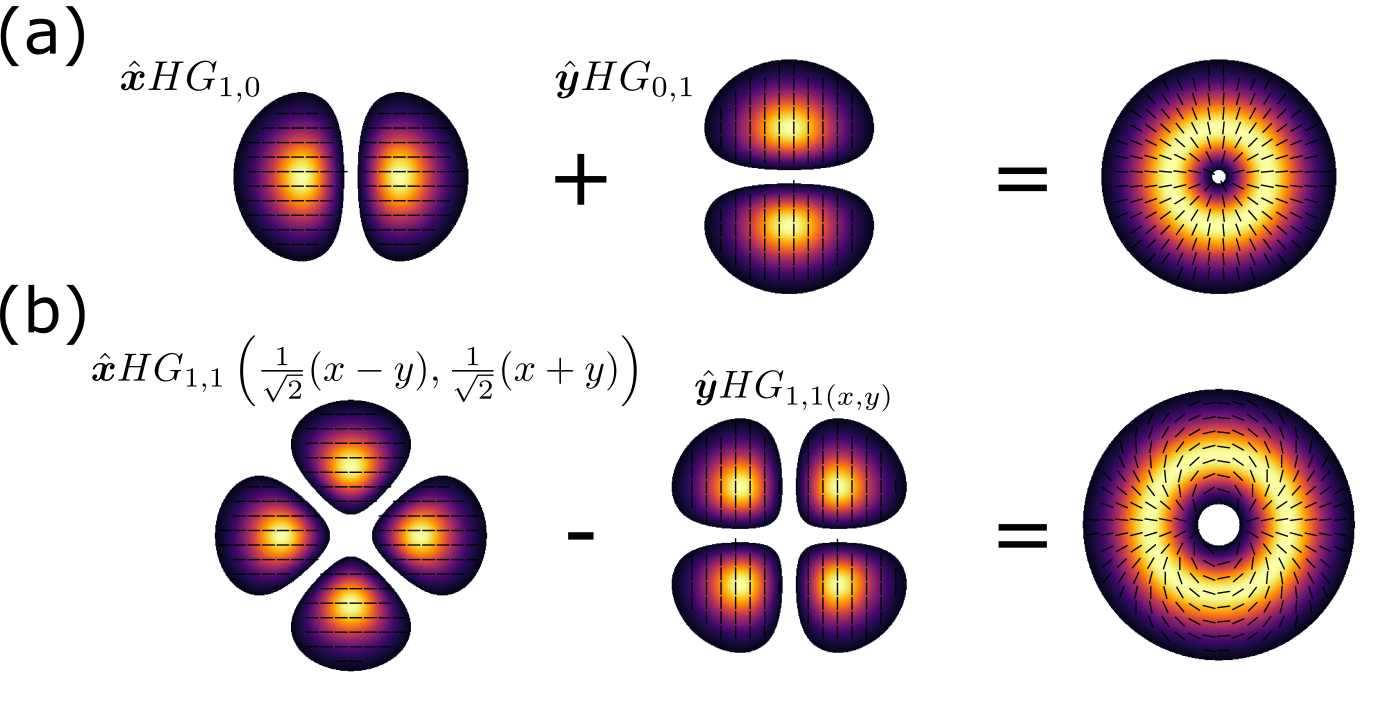}
\caption{\label{fig:HG_Vector_decomp}(a) Visual representation of Hermite-Gaussian beams that are combined to create the $1^{st}$ order vector beam. Hermite-Gaussian modes and vector beams are plotted in intensity with lines representing local polarization. b) The second-order vector beam decomposition.}
\end{figure}

To get a pure superposition of Hermite-Gauss modes, the input would be a linearly polarized beam of the form $(\sqrt{2} r / w)^{p_{vec}} \operatorname{Exp}\left[-r^{2} / w^{2}\right]$. Using this input, we recover the radially-polarized vector beam for $p_{vec} = 1$. A second-order vector beam ($p_{vec} = 2$) is seen to be a superposition of two $HG_{1,1}$ modes as shown in Figure \ref{fig:HG_Vector_decomp}. 
An explicit representation of this second-order vector beam in terms of the Hermite-Gaussian basis is a sum of $HG_{1,1}$ modes, one of which is rotated: 
\begin{eqnarray}
E_{\boldsymbol{V}2} = HG_{1,1}(\frac{1}{\sqrt{2}}(x-y),\frac{1}{\sqrt{2}}(x+y),z)~\hat{\boldsymbol{x}}\nonumber\\ 
    -HG_{1,1}(x,y,z)~\hat{\boldsymbol{y}}.
\end{eqnarray}

The $\hat{\boldsymbol{x}}$ polarization component is spatially rotated 45 degrees to complement the $HG_{1,1}$ mode. Thus, we can see from Fig.\ref{fig:HG_Vector_decomp}b that a linear polarizer can be again used to produce an $HG_{1,1}$ from a second-order vector beam.

By extension, passing a vector beam of order $p_{vec}$ through a linear polarizer will produce a 'necklace' beam with $2 p_{vec}$ lobes of alternating sign. These modes are more conveniently represented as superpositions of Laguerre-Gaussian modes with opposite-sign azimuthal index. 

\section{\label{app:appxB}Analytic calculation of experimentally-produced Hermite-Gaussian beams near focus}

The SWP conversion of an input Gaussian beam to a vector beam is accomplished only by adjusting the local polarization state, producing a polarization singularity at the center of the beam. As the beam propagates away from the SWP, the central dark hole in the beam develops diffractively. The output beam is very close to the desired Hermite-Gaussian mode. By calculating the modal overlap, we find that the radially polarized beam passed through the polarizer has ~93\% overlap with the $HG_{1,0}$ mode when the fundamental Gaussian mode radius is chosen to be $w_0/\sqrt{2}$. 

While it is straightforward to project the beam produced by polarization filtering of the vector beam onto a particular Hermite-Gauss mode, full decomposition into Hermite-Gauss modes to calculate the focused beam is impractical owing to the singularity in the initial field. However, it is possible to calculate the far-field distribution analytically under the Fresnel approximation. For convenience, we start with the vector plate placed at the back focal plane of the lens and use the Fourier transform to the front focal plane, followed by Fresnel propagation to positions near focus. This calculation is most easily done in cylindrical coordinates with the result for the $HG_{1,0}$-like beam
\begin{eqnarray}
        h_{1,0}[\rho, \phi, \zeta]=\boldsymbol{e}^{i k_{0} z_{R} \zeta} \frac{\rho}{(1+i \zeta)^{3 / 2}} e^{-\frac{\rho^{2}}{2(1+i \zeta)}}\nonumber\\ \left(J_{1}\left[i \frac{\rho^{2}}{2(1+i \zeta)}\right]-i J_{0}\left[i \frac{\rho^{2}}{2(1+i \zeta)}\right]\right) \cos [\phi],
\end{eqnarray}
where $\rho=r/w_0$ and $\zeta = z/z_R$ are the normalized radial and longitudinal coordinates, respectively. Similarly, the $HG_{1,1}-$like beam produced with the $2^{nd}-$order S-waveplate gives 
%by transforming the input function $g_{2}[r, \theta]=e^{-r^{2} / w_{\text {in }}{ }^{2}} \cos [2 \theta]$ to obtain 
the field near focus 
\begin{equation}
    h_{1,1}[\rho, \phi, \zeta]=\boldsymbol{e}^{i k_{0}z_R \zeta} \frac{1}{\rho^{2}}\left(1-\left(\frac{\rho^{2}}{1+i \zeta}+1\right) e^{-\frac{\rho^{2}}{1+i \zeta}}\right) \cos [2 \phi].
\end{equation}

\section{\label{app:appxHG11}Comparison between experimental and numerical results for HHG driven by a $HG_{1,1}$ mode}

\begin{figure*}
  \makebox[\textwidth]{
    \includegraphics[width = 0.7\textwidth]{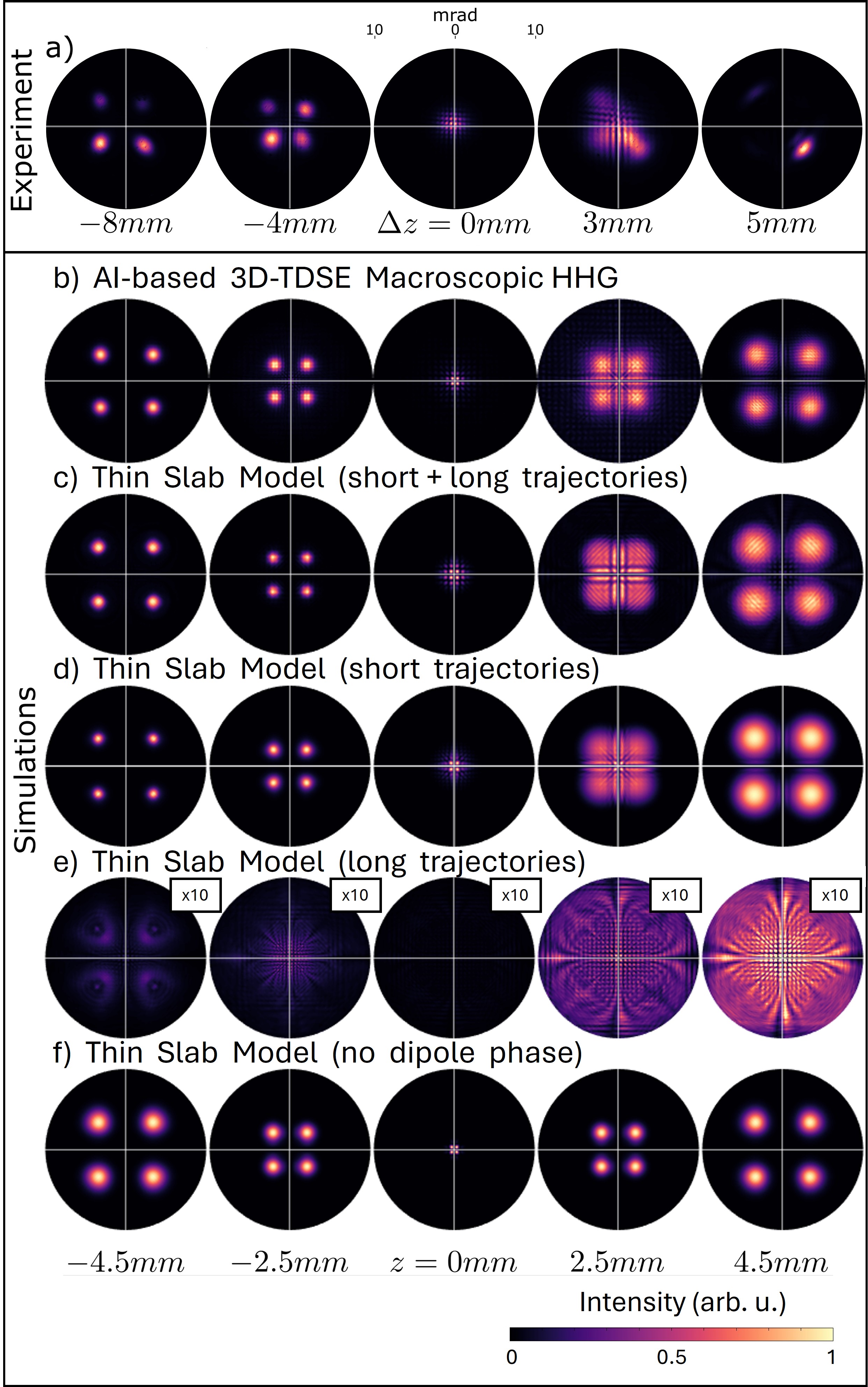} 
  }
   \caption{Theory and experimental comparison of Hermite-Gauss-driven HHG using a $HG_{1,1}$ mode. a) Experimental results of the whole harmonic beam at five positions along the propagation distance. The results are plotted in linear scale, along the far-field divergence. The experimental data is shifted in $\Delta z$ to match the simulation results. b) Far-field intensity distribution using the numerical AI-based 3D-TDSE macroscopic HHG results in section \ref{AI_model}. c) to f) show the corresponding results using the TSM and taking into account c) short and long, d) only short, and e) only long trajectory contributions within the dipole phase term, f) neglecting the dipole phase term. The long trajectory contributions in e) are artificially enhanced by a factor of 10 to make them visible.} The simulations assume an ideal $HG_{1,1}$ with waist parameters matching the experiment. $z$ for the simulation results are absolute and so the center harmonic image was generated at the focus of the ideal $HG_{1,1}$ beam.
	\label{fig:HG11_HHG_Sim}
\end{figure*}

The experimental-theoretical comparison for HHG driven by a $HG_{1,1}$ mode is presented in Fig. \ref{fig:HG11_HHG_Sim}. The first row (a) presents the HHG beam as a function of the relative position between the gas jet position and the focal plane, for five different positions. The second row (b) shows the results form the AI-based 3D TDSE numerical simulations, which are in very good agreement with the experimental results. Main deviations arise from imperfect modal driving beam structure in the experiment. In the third (c) and fourth (d) rows, the theoretical results using the TSM are presented, including or neglecting the dipole phase term, respectively. As it was the case for the $HG_{0,1}$ driving beam, the dipole term is calculated for the so-called short trajectory contributions. Through proper comparison, it can be observed the relevant role in the dipole phase, which breaks the symmetry in the HHG profiles when being generated before or after the focal plane.

\nocite{*}
\bibliography{aipsamp}% Produces the bibliography via BibTeX.

\end{document}